\documentstyle[multicol,eqsecnum,aps]{revtex}

\renewcommand{\narrowtext}{\begin{multicols}{2}
\global\columnwidth20.5pc\noindent}
\renewcommand{\widetext}{\end{multicols}
\global\columnwidth42.5pc}
\begin{document}
\draft
\preprint{30 October 1997}
\title{Elementary Excitations of Heisenberg Ferrimagnetic\\
       Spin Chains}
\author{Shoji Yamamoto}
\address{Department of Physics, Faculty of Science, Okayama University,\\
         Tsushima, Okayama 700, Japan}
\author{S. Brehmer and H.-J. Mikeska}
\address{Institut f\"ur Theoretische Physik, Universit\"at Hannover,
         30167 Hannover, Germany}

\date{Received \hspace{5cm}}
\maketitle
\begin{abstract}
   We numerically investigate elementary excitations of the Heisenberg
alternating-spin chains with two kinds of spins $1$ and $1/2$
antiferromagnetically coupled to each other.
Employing a recently developed efficient Monte Carlo technique as well
as an exact diagonalization method, we verify the spin-wave argument
that the model exhibits two distinct excitations from the ground state
which are gapless and gapped.
The gapless branch shows a quadratic dispersion in the small-momentum
region, which is of ferromagnetic type.
With the intention of elucidating the physical mechanism of both
excitations, we make a perturbation approach from the decoupled-dimer
limit.
The gapless branch is directly related to spin $1$'s, while the
gapped branch originates from cooperation of the two kinds of spins.
\end{abstract}
\pacs{PACS numbers: 75.10.Jm, 05.30.-d, 75.40.Mg, 75.30.Ds}

\narrowtext
\section{Introduction}\label{S:I}

   Extensive efforts have so far been devoted to verifying Haldane's
conjecture \cite{Hald1} that the one-dimensional spin-$S$ Heisenberg
antiferromagnet exhibits qualitatively different properties according
to whether $S$ is integer or half odd integer.
The nontrivial energy gap immediately above the ground state was
precisely estimated using numbers of numerical tools not only in
the spin-$1$ case \cite{Whit1,Sore1,Goli1} but also in the spin-$2$ case
\cite{Scho1,Yama1}, while the valence-bond-solid model \cite{Affl1}
introduced by Affleck, Kennedy, Lieb, and Tasaki significantly
contributed to the understanding of the physical mechanism of 
the so-called
Haldane massive phase.
On the other hand, developing the $O(3)$ nonlinear-$\sigma$-model
quantum field theory \cite{Hald1}, Affleck \cite{Affl2} pointed out that
even integer-spin chains should be critical if a certain interaction is
added to the pure Heisenberg Hamiltonian.
Actually various numerical methods
\cite{Sing1,Kato1,Yama2,Yama3,Tots1,Yaji1,Yama4}
revealed that the spin quantum number is no more the criterion for the
critical behavior in a wider Hamiltonian space.
Recently several authors \cite{Oshi1,Tots2} even suggested the
appearance of the Haldane-gap phases in half-odd-integer-spin chains
with a magnetic field applied.
Thus the low-temperature properties of one-dimensional quantum
antiferromagnets with one kind of spins have more and more been
elucidated.

   In such circumstances, there has appeared brand-new attempts \cite
{Alad1,Vega1,Schl1,Fuji1,Alca1,Kole1,Pati1,Breh1,Fuku1,Nigg1,Ono1,Tone1,Kura1}
to explore the quantum behavior of mixed-spin chains with two kinds of
spins.
These studies are further classified according to their main interests.
Several authors \cite{Alad1,Vega1,Schl1,Fuji1} have been devoting their
efforts to finding quantum integrable Hamiltonians and clarifying their
critical behavior.
Although the models considered are generally complicated, the generic
description of a certain family of Hamiltonians is interesting in
itself and even allows us to guess the essential consequences of
mixed-spin chains.
A distinct attention is directed to mixed-spin chains with the simplest
interaction between the two kinds of spins.
Alternating-spin Heisenberg antiferromagnets with a singlet ground state
\cite{Fuku1,Tone1} again present us the nontrivial gap problem
\cite{Hald1,Affl2}.
Recently Fukui and Kawakami \cite{Fuku1} have made a
nonlinear-$\sigma$-model approach for a few models of this kind and have
discussed a generic criterion for the critical mixed-spin chains.
Their finding may stimulate many theoreticians to numerically investigate
a variety of mixed-spin Hamiltonians and even lead to synthesis of novel
mixed-spin-chain materials.
On the other hand, considering that all the mixed-spin-chain compounds
synthesized so far exhibit a finite ground-state magnetization
\cite{Kahn1}, we take a great interest in alternating-spin Heisenberg
antiferromagnets with ferrimagnetic ground states.
This is the subject we discuss in the present article.

   Let us introduce a Hamiltonian of alternatively aligned two kinds of
spins $S$ and $s$ which are antiferromagnetically coupled to each other:
\begin{equation}
   {\cal H}=J\sum_{j=1}^N
           \left(
            \mbox{\boldmath$S$}_{j} \cdot \mbox{\boldmath$s$}_{j}
           +\delta
            \mbox{\boldmath$s$}_{j} \cdot \mbox{\boldmath$S$}_{j+1}
           \right) \,,
   \label{E:H}
\end{equation}
where
$\mbox{\boldmath$S$}_{j}^2=S(S+1)$,
$\mbox{\boldmath$s$}_{j}^2=s(s+1)$, and
$N$ is the number of unit cells.
The bond alternation $\delta$ has been introduced for a discussion
presented afterwards.
We assume $S>s$ in the following without losing generality.
Because of the non-compensating sublattice magnetizations, this system
exhibits the ferrimagnetism instead of the antiferromagnetism.
Applying the Lieb-Mattis theorem \cite{Lieb1} to the Hamiltonian
(\ref{E:H}), we immediately find $(S-s)N$-fold degenerate ground
states.
The Goldstone theorem \cite{Gold1} further allows us to expect a gapless
excitation from the ferrimagnetic ground state.
Therefore we here take little interest in the simple problem whether
the system is gapped or gapless.
Alcaraz and Malvezzi \cite{Alca1} investigated  the two cases of
$(S,s)=(1,1/2)$ and $(S,s)=(3/2,1/2)$ and actually showed that in both
cases the chain is described in terms of the $c=1$ Gaussian conformal
field theory.
Suggesting that this should be the generic scenario for arbitrary
ferrimagnetic Heisenberg chains, they further predicted the appearance
of quadratic dispersion relations at the Heisenberg points, which means
that the model possesses a ferromagnetic character.
On the other hand, applying the spin-wave theory to the model, 
two groups \cite{Pati1,Breh1} have recently predicted that there exists
a gapped branch of elementary excitations as well as a gapless
ferromagnetic branch.
Their prediction is quite interesting because it implies the
coexistence of the ferromagnetism and the antiferromagnetism in the
ferrimagnets.
This is the motivation for the present study.
Employing a quantum Monte Carlo technique and an exact diagonalization
method, we here calculate energy eigenvalues of the elementary
excitations with the total magnetization
$\sum_{j=1}^N(S_j^z+s_j^z)\equiv M=(S-s)N\mp 1$, which correspond to
the ferromagnetic and the antiferromagnetic branches, respectively.
The numerical results are compared not only with the spin-wave
calculation but also with a perturbation approach from the
decoupled-dimer limit ($\delta=0$).

\section{The Spin-Wave Approach}\label{S:SWA}

   First, we briefly review the spin-wave-theory result
\cite{Pati1,Breh1}, which allows us to have a qualitative view of the
low-energy structure.
We start from a N\'eel state with $M=(S-s)N$, namely, we define the
bosonic operators for the spin deviation in each sublattice as
\begin{equation}
   \left.
   \begin{array}{ccc}
      S_j^+=\sqrt{2S}\,a_j\,,&
      S_j^-=\sqrt{2S}\,a_j^\dagger\,,&
      S_j^z=S-a_j^\dagger a_j\,,\\
      s_j^+=\sqrt{2s}\,b_j^\dagger\,,&
      s_j^-=\sqrt{2s}\,b_j\,,&
      s_j^z=-s+b_j^\dagger b_j\,.
   \end{array}
   \right.
   \label{E:HPBoson}
\end{equation}
In order to obtain the dispersion relations of the spin-wave excitations,
we handle the boson Hamiltonian up to quadratic order.
We define the momentum representation of the bosonic operators as
\begin{equation}
   \left.
   \begin{array}{ccc}
      a_k^\dagger&=&
         \frac{\displaystyle 1}{\displaystyle \sqrt{N}}
         {\displaystyle \sum_{j=1}^N}{\rm e}^{-2{\rm i}ajk}a_j^\dagger\,,\\
      b_k^\dagger&=&
         \frac{\displaystyle 1}{\displaystyle \sqrt{N}}
         {\displaystyle \sum_{j=1}^N}{\rm e}^{ 2{\rm i}ajk}b_j^\dagger\,,\\
   \end{array}
   \right.
   \label{E:Fourier}
\end{equation}
where
$k=\pi l/Na$ ($l=-N/2+1,-N/2+2,\cdots,N/2$) with $a$ being the distance
between two neighboring spins.
We note that here the unit cell is of length $2a$.
Carrying out a Bogoliubov transformation
\begin{equation}
   \left.
   \begin{array}{ccccc}
      \alpha_k&=&
         {\rm e}^{-{\rm i}a\lambda_k/2}{\rm cosh}\theta_k\,a_k
         &+&
         {\rm e}^{ {\rm i}a\lambda_k/2}{\rm sinh}\theta_k\,b_k^\dagger\,,\\
      \beta_k&=&
         {\rm e}^{ {\rm i}a\lambda_k/2}{\rm sinh}\theta_k\,a_k^\dagger
         &+&
         {\rm e}^{-{\rm i}a\lambda_k/2}{\rm cosh}\theta_k\,b_k\,,
   \end{array}
   \right.
   \label{E:Bogol1}
\end{equation}
with
\begin{equation}
   (1+\delta)\tan[(k-\lambda_k)a]-(1-\delta)\tan(ak)=0\,,
   \label{E:Bogol2}
\end{equation}
\begin{eqnarray}
   \tan(2\theta_k)&=&\frac{\displaystyle 2\sqrt{Ss}}
                        {\displaystyle (1+\delta)(S+s)}
                  \nonumber \\
                  &\times&
                   \sqrt{(1+\delta)^2\cos^2(ak)
                        +(1-\delta)^2\sin^2(ak)}\,,
   \label{E:Bogol3}
\end{eqnarray}
we reach the diagonal Hamiltonian,
\begin{equation}
   {\cal H}=E_0
           +\sum_k
            \left(
            \omega_k^-\alpha_k^\dagger\alpha_k
           +\omega_k^+\beta_k^\dagger\beta_k
            \right)\,,
   \label{E:SWH}
\end{equation}
where 
\widetext
\begin{eqnarray}
   E_0&=&-J(1+\delta)SsNJ \nonumber\\
      &+&\frac{\displaystyle J}{\displaystyle 2}
         \sum_k
         \left[
         \sqrt{(1+\delta)^2(S-s)^2+16\delta Ss\sin^2(ak)}-(1+\delta)(S+s)
         \right]\,,
   \label{E:E0}
\end{eqnarray}
\begin{equation}
   \omega_k^\mp=
      \frac{\displaystyle J}{\displaystyle 2}
      \left[
      \sqrt{(1+\delta)^2(S-s)^2+16\delta Ss\sin^2(ak)}\mp(1+\delta)(S-s)
      \right]\,.
   \label{E:omega}
\end{equation}
\narrowtext
Thus the spin-wave approach suggests the coexistence of the
ferromagnetism and the antiferromagnetism in the present system.
In the ferromagnetic branch ($\omega_k^-$), the spin wave reduces the
total magnetization and exhibits a quadratic dispersion relation in
the small-momentum region:
\begin{equation}
   \omega_{k\rightarrow 0}^-
      =\frac{J\delta Ss(2ak)^2}{(1+\delta)(S-s)}\,,
   \label{E:SWFM}
\end{equation}
while in the antiferromagnetic branch ($\omega_k^+$), the spin wave
enhances the total magnetization and have the gapped excitation
spectrum:
\begin{equation}
   \omega_{k\rightarrow 0}^+
      =J(1+\delta)(S-s)+\frac{J\delta Ss(2ak)^2}{(1+\delta)(S-s)}\,.
   \label{E:SWAFM}
\end{equation}
We note that the qualitative character of the model remains unchanged
over the whole region of $\delta$.
It is the purpose in this article to confirm this scenario employing
efficient numerical methods.

\section{A Perturbation Approach}\label{S:PA}

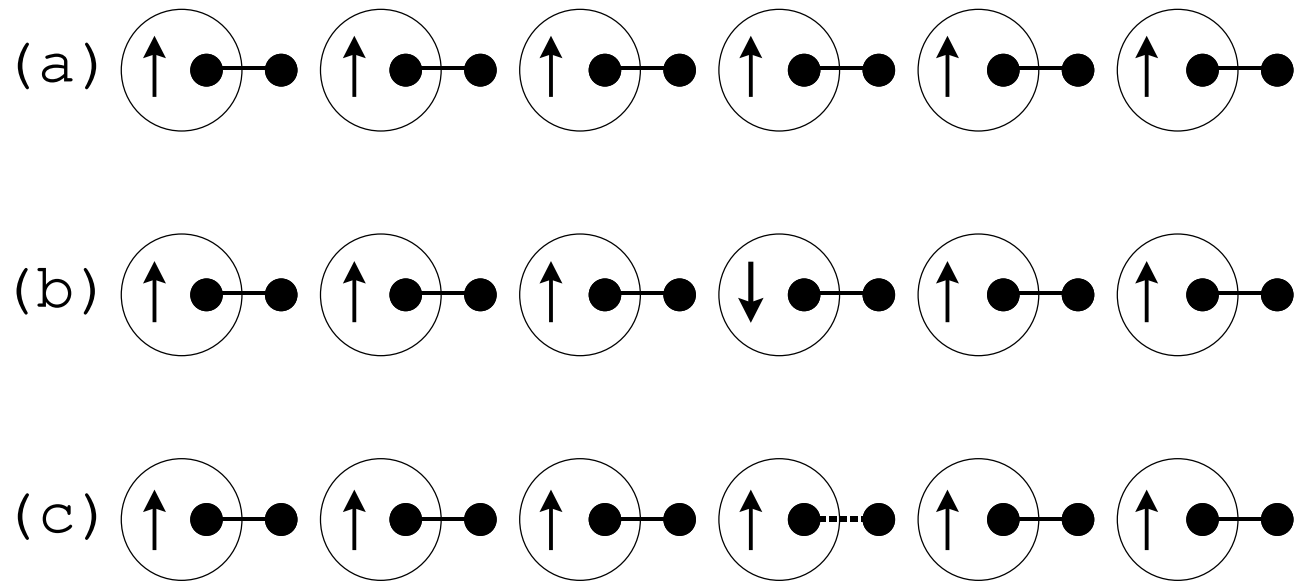
\begin{figure}
\caption{Schematic representations of the ground state (a), the elementary
         excitation in the subspace of $M=N/2-1$ (b), and the elementary
         excitation in the subspace of $M=N/2+1$ (c) of the ferrimagnetic
         chain with spin $1$ and spin $1/2$ in the decoupled-dimer limit.
         The arrow (the bullet symbol) denotes a spin $1/2$ with its
         fixed (unfixed) projection value.
         The solid (broken) segment is a singlet (triplet) pair.
         The circle represents an operation of constructing a spin $1$ by
         symmetrizing the two spin $1/2$'s inside.}
\label{F:illust1}
\end{figure}

   Second, prior to the numerical approach, let us carry out a
perturbation calculation with the intention of elucidating the nature of
the elementary excitations.
In the decoupled-dimer limit, we can easily find the low-lying
eigenstates.
Figures \ref{F:illust1}(a), \ref{F:illust1}(b), and \ref{F:illust1}(c)
represent, respectively, the ground state, $|\Psi\rangle$,
the ferromagnetic excitation at an arbitrary unit cell $j$,
$|\Psi_j^\downarrow\rangle$, and the antiferromagnetic excitation at
an arbitrary unit cell $j$, $|\Psi_j^+\rangle$, of the Hamiltonian
(\ref{E:H}) with $(S,s)=(1,1/2)$ at $\delta=0$.
When we turn on the exchange interaction between the dimers, the
localized excitations can hop to neighboring unit cells with an
amplitude proportional to $\delta$.
We take account of this effect using the degenerate perturbation.
We introduce a representation of matrix-product type as
\begin{eqnarray}
   |\Psi\rangle\ 
      &=&g_1^{\uparrow s}\otimes h_1^s
         \otimes\cdots\otimes
         g_N^{\uparrow s}\otimes h_N^s
         \,, \\
   |\Psi_j^\downarrow\rangle
      &=&g_1^{\uparrow s}\otimes h_1^s
         \otimes\cdots\otimes
         g_j^{\downarrow s}\otimes h_j^s
         \otimes\cdots\otimes
         g_N^{\uparrow s}\otimes h_N^s
         \,, \\
   |\Psi_j^+\rangle
      &=&g_1^{\uparrow s}\otimes h_1^s
         \otimes\cdots\otimes
         g_j^{\uparrow +}\otimes h_j^+
         \otimes\cdots\otimes
         g_N^{\uparrow s}\otimes h_N^s
         \,,
   \label{E:Psi}
\end{eqnarray}
where
\begin{equation}
   \left.
   \begin{array}{ccc}
   g_j^{\uparrow s}&=&
      \left[
      \begin{array}{cc}
         |0\rangle_j & \sqrt{2}|+\rangle_j
      \end{array}
      \right]\,,\\
   g_j^{\downarrow s}&=&
      \left[
      \begin{array}{cc}
         \sqrt{2}|-\rangle_j & |0\rangle_j
      \end{array}
      \right]\,,\\
   g_j^{\uparrow +}&=&
      \left[
      \begin{array}{cc}
         -\sqrt{2}|+\rangle_j & 0
      \end{array}
      \right]\,,
   \end{array}
   \right.
   \label{E:g}
\end{equation}
\begin{equation}
   h_j^{s}=
      \left[
      \begin{array}{c}
         -|\uparrow  \rangle_j \\
          |\downarrow\rangle_j
      \end{array}
      \right]\,,\ \
   h_j^{\uparrow +}=
      \left[
      \begin{array}{c}
         -|\uparrow\rangle_j \\
          0
      \end{array}
      \right]\,,
   \label{E:h}
\end{equation}
with $|\pm\rangle_j$, $|0\rangle_j$ being the $S_j^z$-eigenstates and
$|\uparrow\rangle_j$, $|\downarrow\rangle_j$ the $s_j^z$-eigenstates.
Now the dispersion relations of the eigenstates with $M=N/2\mp 1$,
$\omega_k^\mp$, are calculated as
\begin{eqnarray}
   \omega_k^-
      &=&\frac{\langle\Psi_k^\downarrow|{\cal H}|\Psi_k^\downarrow\rangle}
              {\langle\Psi_k^\downarrow|         \Psi_k^\downarrow\rangle}
        -E_{\rm G}
       \nonumber \\
      &=&\frac{4}{9}J\delta[1-\cos(2ak)]
        +O(\delta^2)\,, \\
   \omega_k^+
      &=&\frac{\langle\Psi_k^+         |{\cal H}|\Psi_k^+         \rangle}
              {\langle\Psi_k^+         |         \Psi_k^+         \rangle}
        -E_{\rm G}
       \nonumber \\
      &=&\frac{3}{2}J+\frac{1}{18}J\delta[7-12\cos(2ak)]
        +O(\delta^2)\,,
   \label{E:omega1}
\end{eqnarray}
where
$|\Psi_k^\downarrow\rangle
 =N^{-1/2}\sum_{j=1}^N{\rm e}^{-2{\rm i}ajk}
  |\Psi_j^\downarrow\rangle$ and
$|\Psi_k^+\rangle
 =N^{-1/2}\sum_{j=1}^N{\rm e}^{-2{\rm i}ajk}
  |\Psi_j^+\rangle$
are the Fourier transforms of
$|\Psi_j^\downarrow\rangle$ and $|\Psi_j^+\rangle$, respectively, and
$E_{\rm G}=\langle\Psi|{\cal H}|\Psi\rangle
          /\langle\Psi|         \Psi\rangle
          =-J(1+\delta/9)N$
is the ground-state energy within the first order of $\delta$.
Although the Heisenberg point ($\delta=1$) is far from the
decoupled-dimer limit, it is interesting enough that the qualitative
characters of both branches remain unchanged as $\delta$ increases:
The ferromagnetic branch $\omega_k^-$ is gapless and proportional to
$k^2$ in the small-$k$ region, while the antiferromagnetic branch
$\omega_k^+$ is gapped.

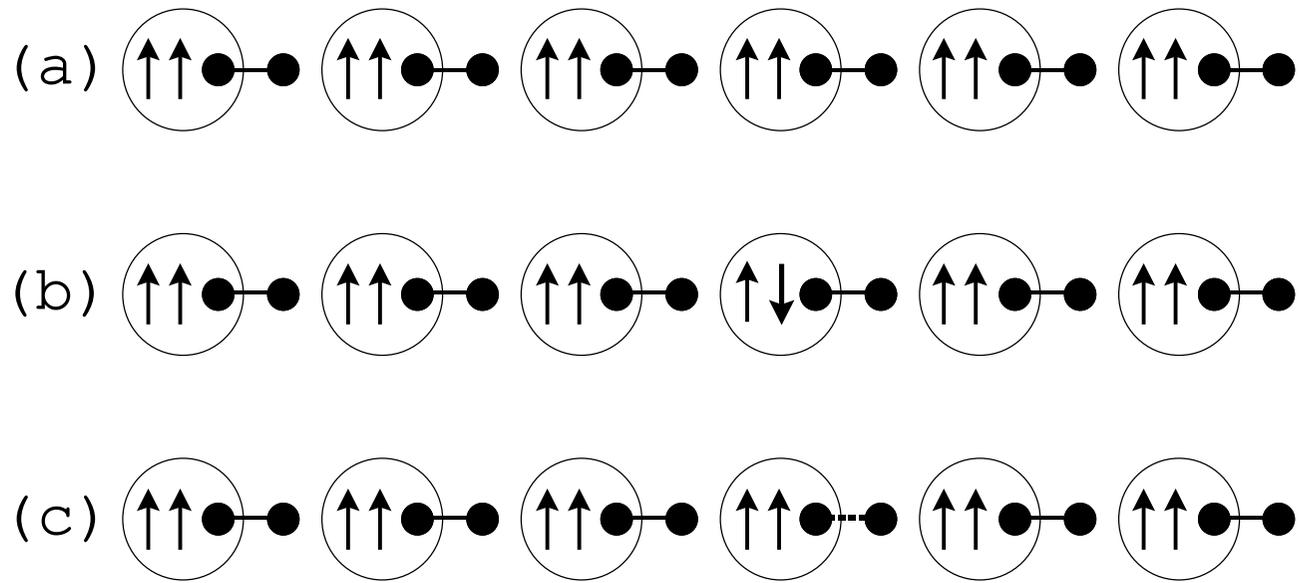
\begin{figure}
\caption{Schematic representations of the ground state (a), the elementary
         excitation in the subspace of $M=N/2-1$ (b), and the elementary
         excitation in the subspace of $M=N/2+1$ (c) of the ferrimagnetic
         chain with spin $3/2$ and spin $1/2$ in the decoupled-dimer limit.
         The notation is the same as one in Fig. 1 except
         for the circle representing an operation of constructing a spin
         $3/2$ by symmetrizing the three spin $1/2$'s inside.}
\label{F:illust2}
\end{figure}

   For an arbitrary combination of $(S,s)$, the similar argument can
be developed and qualitatively the same result is obtained.
For example, in the case of $(S,s)=(3/2,1/2)$, we find the dispersion
relations
\begin{eqnarray}
   \omega_k^-
      &=&\frac{5}{8}J\delta[1-\cos(2ak)]+O(\delta^2)\,,\\
   \omega_k^+
      &=&2J+\frac{1}{8}J\delta[7-6\cos(2ak)]+O(\delta^2)\,,
   \label{E:omega2}
\end{eqnarray}
with $E_{\rm G}=-(5/4)J(1+\delta/4)N$, considering the elementary
excitations shown in Fig. \ref{F:illust2}.
Thus we are more and more convinced that the scenario of the
low-energy structure should be valid for an arbitrary Heisenberg
ferrimagnet.
Actually, recent numerical studies \cite{Alca1,Pati1} on the present
Hamiltonian reported that the low-temperature properties are
essentially the same regardless of the values of $S$ and $s$ as long
as they differ from each other.
Alcaraz and Malvezzi \cite{Alca1} found that the model with exchange
anisotropy is described in terms of the Gaussian critical theory in
both cases of $(S,s)=(1,1/2)$ and $(S,s)=(3/2,1/2)$.
In such circumstances, we restrict our numerical investigation to the
case of $(S,s)=(1,1/2)$.

\section{Numerical Procedure}\label{S:NP}

   In order to calculate the low-lying eigenvalues of the model, we
here use two numerical tools, which possess, respectively, both
advantageous and weak points of their own and are complementary to each
other.

   One is the exact-diagonalization method employing the Lanczos
algorithm.
Calculation of the energy levels reduces to the diagonalization of the
$6^N\times 6^N$ matrix representing the Hamiltonian.
In constructing basic states, we use direct products of the single-spin
states indicated by the projection values.
The present Hamiltonian commutes with the total magnetization $M$ and
therefore splits into $3N+1$ blocks labeled by $M$ (on the assumption
that $N$ is even).
Since we perform the calculation under the periodic boundary condition,
each eigenvalue is further classified by its total wave number $k$.

   We treat the chains of $N=8,10,12$, where we restrict the
calculation to the lowest energy level in each subspace.
Since our main interest is to reveal the nature of the elementary
excitations, we investigate the subspaces of $M=N/2$ and $M=N/2\mp 1$
for all the chain lengths we treat.
We calculate the subspaces of $M\leq N/2-2$ as well for $N=8,10$ in
order to further elucidate the ferromagnetic nature of the model.
Due to the two kinds of spins, the construction of the basic states
is somewhat complicated.
The base dimension reaches ten million in the case of $N=12$ and
$M=N/2-1$.

   The chain length we can reach with the diagonalization method is
much smaller than one with a Monte Carlo technique.
However, having in mind the small correlation length of the present
system \cite{Pati1,Breh1}, the diagonalization result is fruitful
enough to obtain a general view of the low-energy structure.
Actually, the ground-state energy for $N=12$ coincides with the
$N\rightarrow\infty$ extrapolated value within the first five digits.
We further note that even the ground-state energy for $N=8$ is so
close to the thermodynamic-limit value as to show the coincidence
within the first three digits.
This fact is quite helpful in estimating the dispersion relations in
the long-chain limit, although the eigenvalues for different chain
lengths should be distinguished.

   The other approach is based on a quantum Monte Carlo technique
\cite{Yama3,Yama5,Yama6,Yama7} which one of the authors has
recently developed.
The idea is summarized as extracting the lower edge of the excitation
spectrum from imaginary-time quantum Monte Carlo data at a low enough
temperature.
The imaginary-time correlation function $S(q,\tau)$ is generally
defined as
\begin{equation}
   S(q,\tau)=\left\langle
   {\rm e}^{{\cal H}\tau}O_q^z{\rm e}^{-{\cal H}\tau}O_{-q}^z
             \right\rangle\,,
   \label{E:Sqtdef}
\end{equation}
where $O_q=N^{-1}\sum_{j=1}^N O_j{\rm e}^{{\rm i}qja_0}$
is the Fourier transform of an arbitrary local operator $O_j$ with
$a_0$ being the length of the unit cell, and
$\left\langle A\right\rangle \equiv
 {\rm Tr}[{\rm e}^{-\beta{\cal H}} A]/
 {\rm Tr}[{\rm e}^{-\beta{\cal H}}]$
denotes the canonical average at a given temperature
$\beta^{-1}=k_{\rm B}T$.
While $S(q,\tau)$ as a function of $\tau$ generally exhibits a
complicated multi-exponential decay, it may efficiently be evaluated
at a sufficiently low temperature as \cite{Yama5}
\begin{equation}
   S(q,\tau)=\sum_{l}
             \left\vert \langle 1;k_0 \vert
             S_q^z
             \vert l;k_0+q \rangle \right\vert^2
             {\rm e}^{-\tau\left[E_{l}(k_0+q)-E_1(k_0)\right]} \,,
   \label{E:SqtLT}
\end{equation}
where
$\vert l;k\rangle$ $(l=1,2,\cdots)$ and
$E_{l}(k)$ ($E_1(k) \leq E_2(k) \leq \cdots$)
are the $l$th eigenvector and eigenvalue of the Hamiltonian in the
$k$-momentum space, and $k_0$ is the momentum at which the
lowest-energy state in the subspace is located.
Now it is reasonable to approximate $E_1(k_0+q)-E_1(k_0)$ by the  
slope
$-\partial{\rm ln}[S(q,\tau)]/\partial\tau$ in the large-$\tau$  
region
satisfying
\begin{equation}
   \tau[E_2(k_0+q)-E_1(k_0+q)]\gg
   {\rm ln}
   \frac{\vert\langle 1;k_0 \vert S_q^z\vert n;k_0+q\rangle\vert^2}
        {\vert\langle 1;k_0 \vert S_q^z\vert 1;k_0+q\rangle\vert^2}\,,
   \label{E:cond}
\end{equation}
for an arbitrary $n$.

   In case the lower edge of the spectrum is separated from the upper
bands or continuum by a finite gap, or in case of its spectral weights
$|\langle 1;k_0|O_q^z|1;k_0+q\rangle |^2$
being relatively large, the inequality (\ref{E:cond}) is well justified
even at small $\tau$'s and a logarithmic plot of $S(q,\tau)$ is
expected to exhibit fine linearity in a wide region of $\tau$.
Actually, for single-spin Heisenberg chains with an arbitrary spin
quantum number or a certain bond alternation, it was shown
\cite{Yama3,Yama5,Yama6,Yama7} that the $\tau$-dependence of
$S(q,\tau)$ is essentially approximated by a single exponent at each
momentum $q$ at a sufficiently low temperature.

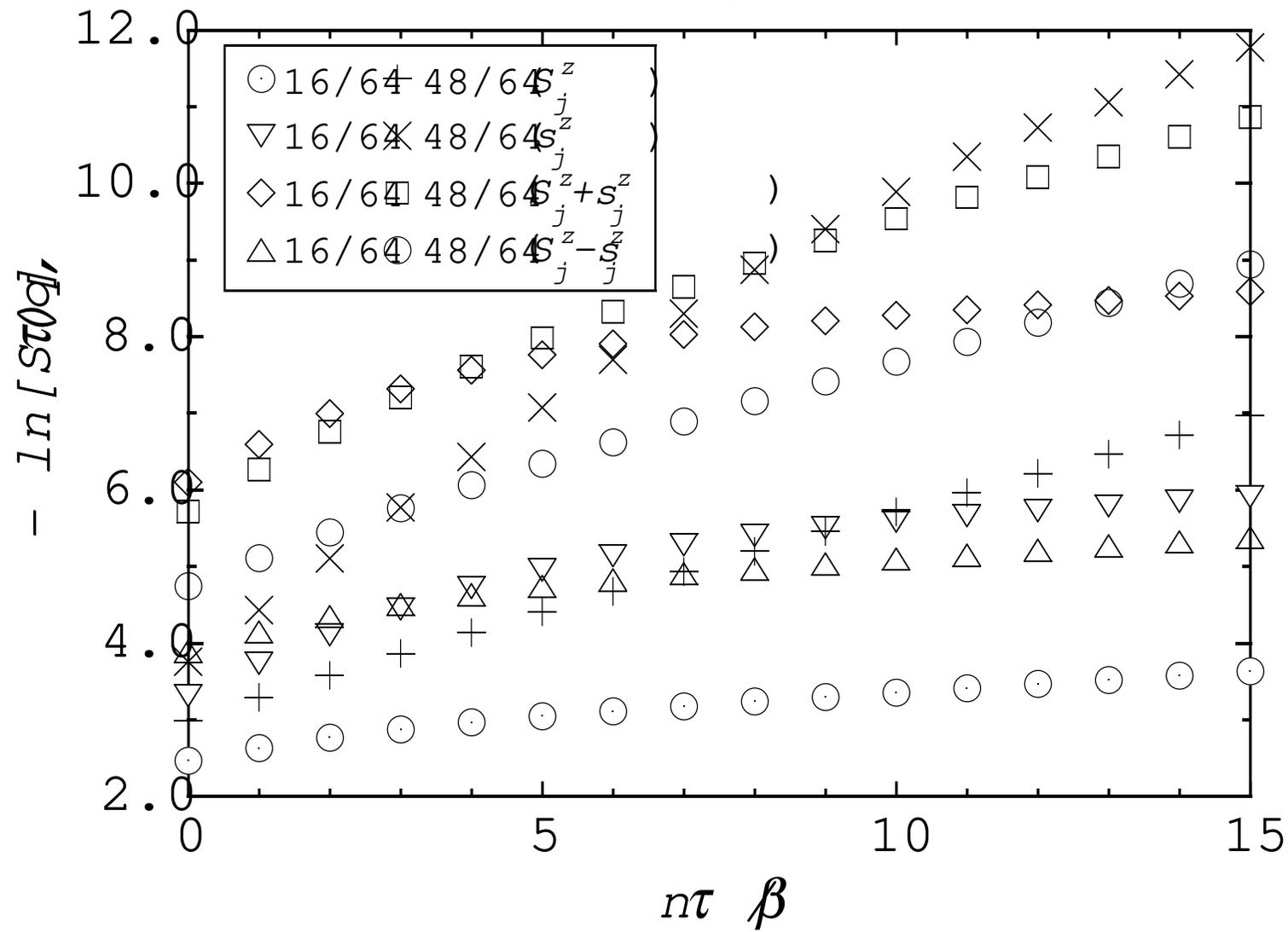
\begin{figure}
\caption{Logarithmic plots of $S(q,\tau)$ versus the imaginary time $\tau$
         at several choices of $O_j$ taking a few values of momentum $q$
         for the Heisenberg ferrimagnetic chain of $N=32$ with
         $(S,s)=(1,1/2)$ and $\delta=1$:
         $\bigodot$         ($2aq/\pi=16/64$),
         $+$                ($2aq/\pi=48/64$)
         with $O_j=S_j^z$;
         $\bigtriangledown$ ($2aq/\pi=16/64$),
         $\times$           ($2aq/\pi=48/64$)
         with $O_j=s_j^z$;
         $\Diamond$         ($2aq/\pi=16/64$),
         $\Box$             ($2aq/\pi=48/64$)
         with $O_j=S_j^z+s_j^z$;
         $\bigtriangleup$   ($2aq/\pi=16/64$),
         $\bigcirc$         ($2aq/\pi=48/64$)
         with $O_j=S_j^z-s_j^z$.
         The numerical uncertainty is all within the size of the
         symbols.}
\label{F:lnSqt}
\end{figure}

   Here, due to the two kinds of spins in a chain, $O_j$ is not
uniquely defined.
We show in Fig. \ref{F:lnSqt} logarithmic plots of $S(q,\tau)$ as a
function of $\tau$ setting several operators for $O_j$.
We have set $(\beta J)^{-1}$ and $n$ equal to $0.02$ and $200$, which
are, respectively, low and large enough \cite{Yama7} to remove the
finite-temperature effect and the finite-$n$ effect.
In order to estimate each $S(q,\tau)$, we have carried out a few
million Monte Carlo steps spending several days on a supercomputer
or a few weeks on a fast workstation.
Energy difference between the ground state and the lowest state with an
arbitrary $q$ is obtained through $S(q,\tau)$ calculated in the
subspace of $M=0$.

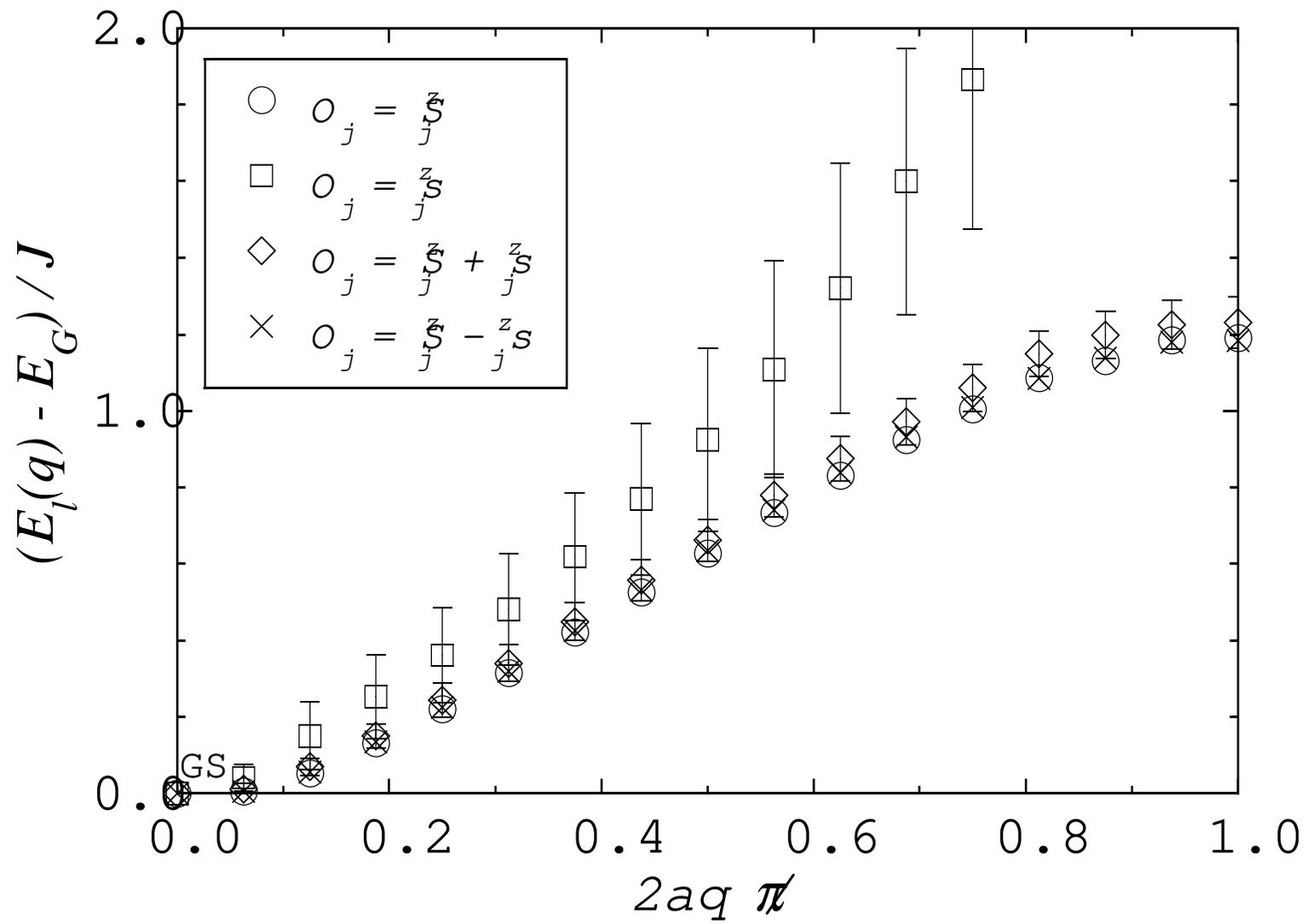
\begin{figure}
\caption{Quantum Monte Carlo estimates of excitation energies as a function
         of $q$ for the chain of $N=32$.
         Here $\bigcirc$, $\Box$, $\Diamond$, and $\times$ have,
         respectively, been obtained from $S(q,\tau)$'s with $O_j=S_j^z$,
         $O_j=s_j^z$, $O_j=S_j^z+s_j^z$, and $O_j=S_j^z-s_j^z$ at the
         subspace of $M=0$.
         The error bars are attached to the data obtained with $O_j=s_j^z$
         and $O_j=S_j^z+s_j^z$.
         The numerical uncertainty of the rest of the data is all within
         the size of the symbols.
         GS denotes the ground state.}
\label{F:dspQMC}
\end{figure}

   In Fig. \ref{F:dspQMC} we plot the excitation energies as a function
of $q$ obtained by estimating the slope
$-\partial{\rm ln}[S(q,\tau)]/\partial\tau$
in the largest-$\tau$ region available.
In the case of $O_j=s_j^z$, the multi-exponential behavior of
$S(q,\tau)$ is remarkable even in the large-$\tau$ region and thus
prevents us from precisely estimating the energy eigenvalues.
In all the other cases, we obtain useful data within the numerical
precision.
With the present data, we can at least conclude that spin-$1/2$'s do
not have much effect on the lowest-lying excitations, which suggests
that the elementary excitations of ferromagnetic nature may
qualitatively be described by the simple picture shown in Fig.
\ref{F:illust1}(b) even at the Heisenberg point $\delta=1$.
While $O_j=S_j^z+s_j^z$ brings somewhat higher energies than
$O_j=S_j^z$ and $O_j=S_j^z-s_j^z$, we find no difference beyond the
numerical accuracy between the cases of $O_j=S_j^z$ and
$O_j=S_j^z-s_j^z$.
It will be shown in the next section that the thus-obtained
lowest-lying energy eigenvalues are in good agreement with the
exact-diagonalization result.
This method is applicable to rather long chains but is not
successful in obtaining the higher-lying eigenvalues except for the
special fortunate cases \cite{Yama5}.
Thus the diagonalization technique is necessary and useful for the
antiferromagnetic branch with $M>N/2$ even though it is inferior to
the Monte Carlo method in treating long chains.

\section{Numerical Results and Discussion}\label{S:NRD}

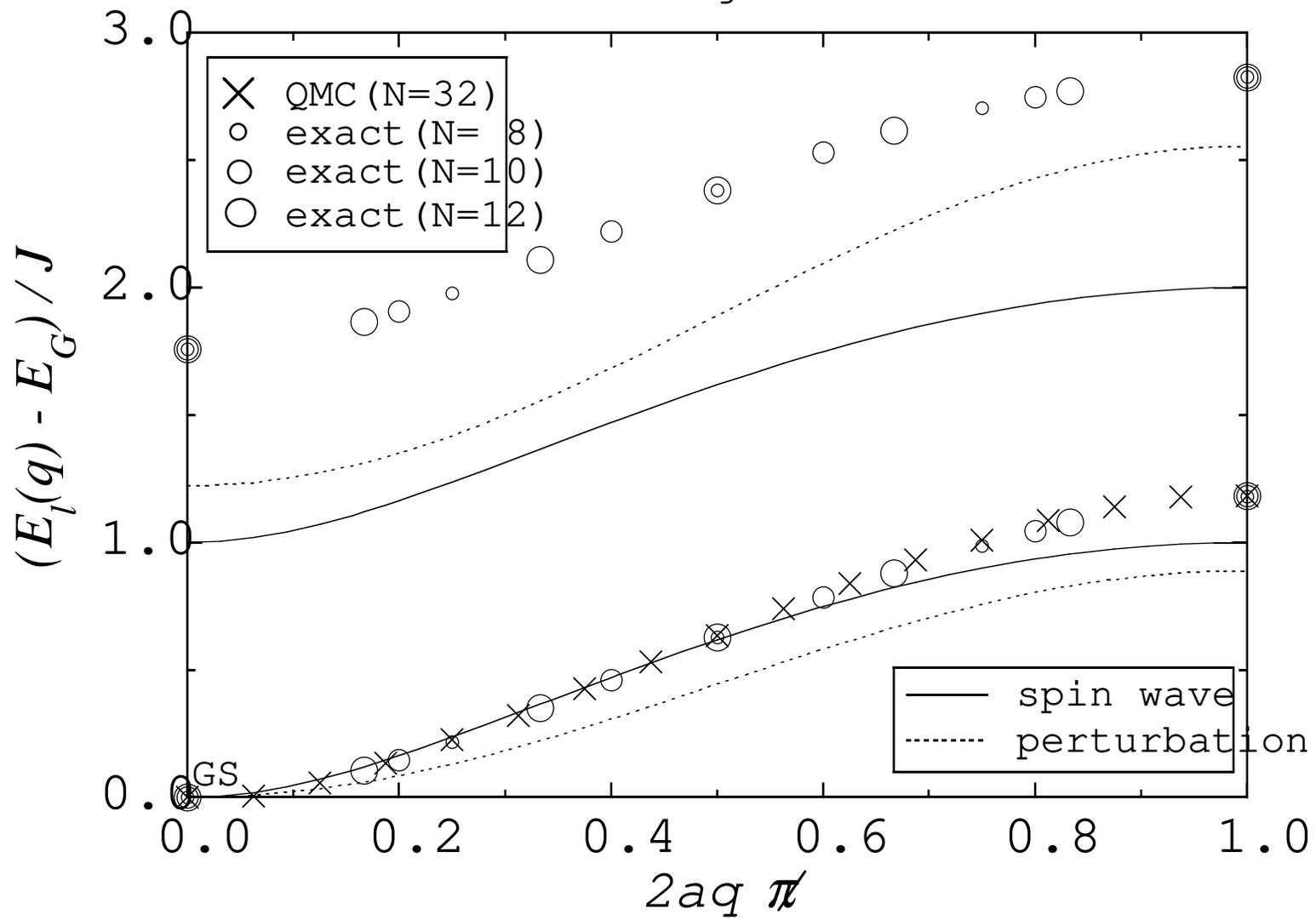
\begin{figure}
\caption{Quantum Monte Carlo and exact-diagonalization calculations of
         the lowest energies as a function of momentum in the subspaces
         of $M=N/2\mp 1$ for the Heisenberg ferrimagnetic chain of
         $N=32$ with spin $1$ and spin $1/2$.
         The numerical uncertainty is all within the size of the
         symbols.
         The results of the spin-wave theory and the first-order
         perturbation from the decoupled-dimer limit are also shown by
         solid and broken lines, respectively.
         GS denotes the ground state.}
\label{F:dspall}
\end{figure}

   We plot in Fig. \ref{F:dspall} the quantum Monte Carlo and the
exact-diagonalization calculations of the excitation energies as a
function of momentum $q$ for the chain without bond alternation, where
the results of the spin-wave theory and the first-order perturbation
from the decoupled-dimer limit are also shown.
The lower band is the lower edge of the excitation spectrum and
consists of the lowest-lying eigenvalues with $M=N/2-1$.
It exhibits a quadratic dispersion at small $q$'s as was expected.
The upper band consists of the lowest-lying eigenvalues with
$M=N/2+1$ and is separated from the ground state by a finite gap.
It is the scenario predicted by the analytic approaches that we here
observe.
The quantum Monte Carlo finding is in good agreement with the
diagonalization result.
The diagonalization calculation indicates that the chain-length
dependence of the dispersion relation is quite weak even in the
vicinity of the zone boundary and the zone center, which is consistent
with the extremely small correlation length \cite{Pati1,Breh1}.

   In order to perform a quantitative comparison between the numerical
findings and the analytic results, let us consider the curvature of the
dispersion, $v$, which is defined by
\begin{equation}
   \omega_k^-=v(2ak)^2\,.
   \label{E:v}
\end{equation}
The quantum Monte Carlo method, the spin wave calculation, and the
perturbation approach, respectively, give
$v^{\rm QMC}/J=0.37(1)$, $v^{\rm SW}/J=1/2$, and $v^{\rm pert}/J=2/9$.
The spin-wave theory overestimates the true value, while the
perturbation approach underestimates that.
We have made an attempt to obtain another estimate of $v$ using the
exact-diagonalization result.
Although the Lanczos algorithm results in much more precise raw data
than the Monte Carlo technique, yet the attempt was not so
successful as one using the Monte Carlo data because of the lack of
data points.
However, we have confirmed that the diagonalization estimate of $v$
possibly coincides with $v^{\rm QMC}$ within the numerical accuracy.
We note that the spin-wave estimate $v^{\rm SW}$ accords with one for
the Heisenberg ferromagnet of spin $1/2$ in the unit of the unit-cell
length being equal to unity.
Furthermore the calculation of the lowest levels in the subspace of
$M= N/2 -2$ results in energy eigenvalues bellow the two-magnon
continuum.
The obtained state should be a two-magnon bound state \cite{Matt1}.
All these facts again emphasize the ferromagnetic aspect of the present
model.
However, in contrast with the ferromagnet, the true value $v^{\rm QMC}$
is reduced from $v^{\rm SW}$ due to the quantum effect.
Thus the elementary excitations in the sector of $M<N/2$ can be regarded
as spin waves modified by quantum fluctuations.

   For the antiferromagnetic branch, on the other hand, both
spin-wave analysis and perturbation calculation are not so
successful as ones for the ferromagnetic branch.
Considering that quantum effects are, in general, much more remarkable
in antiferromagnets than in ferromagnets, the most naive picture of the
elementary excitations illustrated in Fig. \ref{F:illust1} and Fig.
\ref{F:illust2} may have to be significantly modified for the
antiferromagnetic branch.

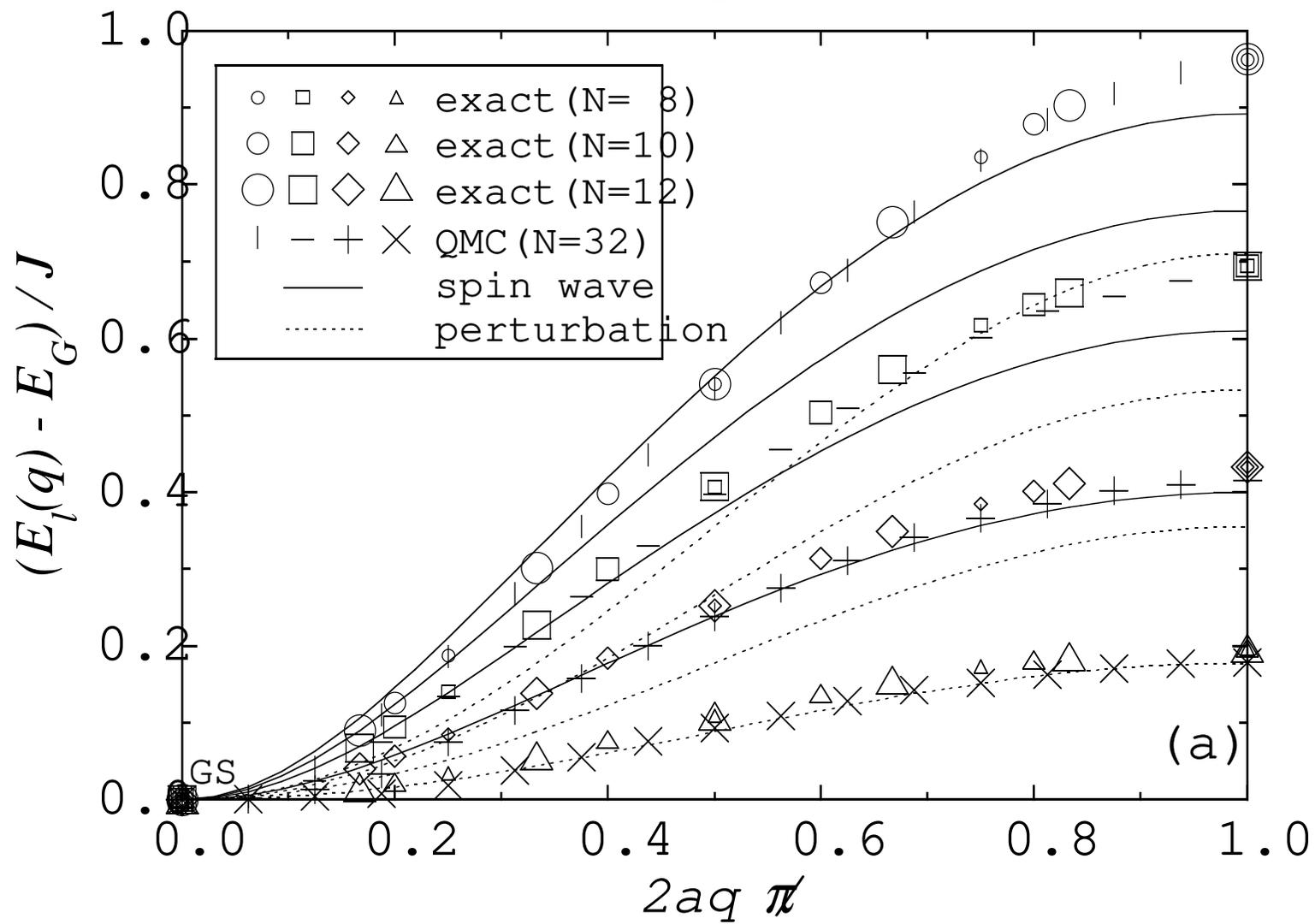
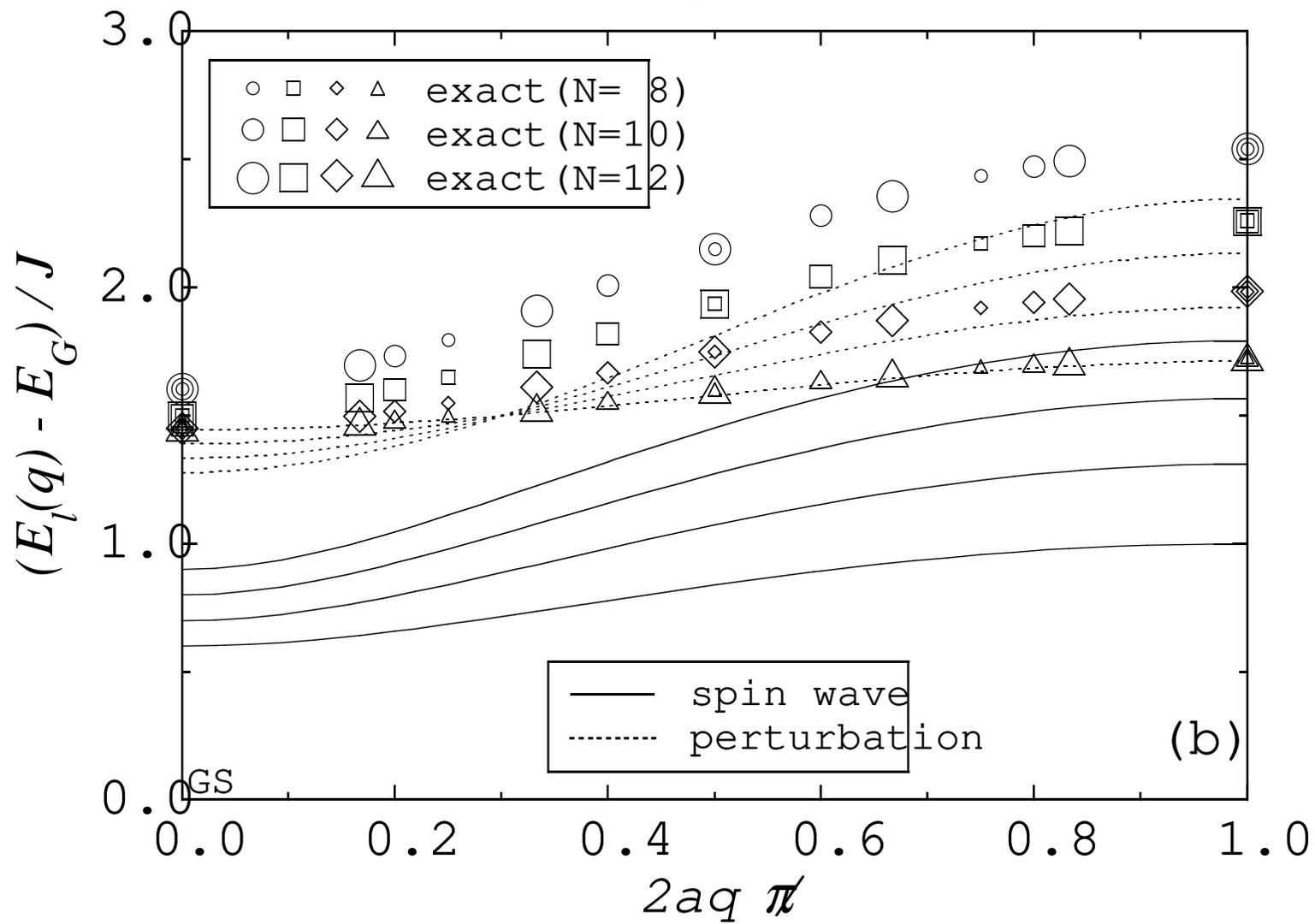
\begin{figure}
\caption{The lowest energies as a function of momentum in the subspaces
         of $M=N/2-1$ (a) and $M=N/2+1$ (b) for the bond-alternating
         Heisenberg ferrimagnetic chains of $N=32$ with spin $1$ and
         spin $1/2$.
         Here $\bigcirc$ ($|$), $\Box$ ($-$), $\Diamond$ ($+$), and
         $\bigtriangleup$ ($\times$) represent the exact-diagonalization
         (quantum Monte Carlo) estimates at $\delta=0.8$, $\delta=0.6$,
         $\delta=0.4$, and $\delta=0.2$, respectively.
         The numerical uncertainty is all within the size of the
         symbols.
         The results of the spin-wave theory and the first-order
         perturbation from the decoupled-dimer limit are also shown by
         solid and broken lines, respectively, where the values of
         $\delta$ are $0.8$, $0.6$, $0.4$, and $0.2$ from top to bottom
         except for the perturbation result for the antiferromagnetic
         branch in the small-$q$ region with the values of $\delta$
         being $0.2$, $0.4$, $0.6$, and $0.8$ from top to bottom.
         GS denotes the ground state.}
\label{F:dspdelta}
\end{figure}

   Finally in this section, we show in Fig. \ref{F:dspdelta} the
calculation for the chains with bond alternation.
Although the quantum Monte Carlo calculation is generally in good
agreement with the diagonalization result, the agreement seems to be
somewhat poorer in the small-$\delta$ region.
This is convincing keeping in mind that the decrease of $\delta$ may
cause the freezing of the spin configuration in Monte Carlo sampling
and therefore a huge number of Monte Carlo steps are needed to refine
the data accuracy in the small-$\delta$ region.
It is needless to say that the perturbation calculation is more
justified in the small-$\delta$ region.
As the model approaches the decoupled-dimer limit, both ferromagnetic
and antiferromagnetic bands become flatter and approach $0$ and
$(3/2)J$, respectively.
The spin-wave result is fairly good in the ferromagnetic branch but
relatively poor in the antiferromagnetic branch.
This is convincing considering that the spin wave correctly describes
the low-lying excitations of the ferromagnets, while it is valid at most
qualitatively for the antiferromagnets.

\section{Summary}\label{S:S}

   We have investigated the low-energy structure of the ferrimagnetic
alternating-spin chains with spin $1$ and spin $1/2$.
Motivated by the spin-wave analysis of the model, we have mainly
calculated the eigenvalues with an arbitrary momentum of the one-magnon
states, namely, the lowest-lying eigenvalues in the subspaces of
$M=N/2\mp 1$.
The chain-length dependence of the dispersion relations is extremely
weak, which is consistent with the considerably small correlation length
of the system, $\xi<2a$ \cite{Pati1,Breh1}.
The qualitative character of the model remains unchanged under the
existence of bond alternation.
The ferromagnetic branch is gapless and shows a quadratic dispersion in
the small-momentum region, while the antiferromagnetic branch is
separated from the ground state by a finite gap $\Delta$.
$\Delta/J$ in the thermodynamic limit is estimated to be $1.75914(1)$ at
the Heisenberg point $\delta=1$.
We made an attempt to understand the low-lying excitations through the
first-order perturbation calculation from the decoupled-dimer limit.
The ferromagnetic excitations are more or less dominated by spin $1$'s,
whereas the mechanism of the antiferromagnetic excitations was less
revealed.
Quantum Monte Carlo snapshots \cite{Yama8} may help us to inquire
further into the antiferromagnetic fluctuations.

   We have further carried out the diagonalization calculation in the
subspace of $M=N/2-2$ and found the two-magnon bound state bellow
the continuum.
This fact emphasizes the ferromagnetic aspect of the sector of $M<N/2$
in the ferrimagnet.
We expect that the Heisenberg ferrimagnet behaves like a ferromagnet
at low enough temperatures, while its antiferromagnetic aspect may
appear at $k_{\rm B}T\agt\Delta$.

\acknowledgments

   A part of this work was carried out while S.Y. was staying at
Hannover Institute for Theoretical Physics.
S.Y. gratefully acknowledges the hospitality of the institute at that
time.
The authors would like to thank U. Neugebauer for his great help in
coding the Lanczos algorithm.
They are further grateful to T. Tonegawa, T. Fukui, and S. K. Pati for
their useful comments.
This work was supported by the German Federal Ministry for Research
and Technology (BMBF) under the Contract 03-MI4HAN-8, by the Japanese
Ministry of Education, Science, and Culture through the Grant-in-Aid
09740286, and by a Grant-in-Aid from the Okayama Foundation for Science
and Technology.
Most of the numerical computation was done using the facility of the
Zuse Computing Center, Berlin and one of the Supercomputer Center,
Institute for Solid State Physics, University of Tokyo.

\widetext
\end{document}